\documentclass[10pt, article, nofootinbib]{revtex4}
\usepackage{graphicx}
\usepackage{booktabs}
\usepackage{amssymb}
\usepackage{multirow}
\usepackage{siunitx}
\usepackage{hyperref}
\usepackage[utf8]{inputenc} 

\begin{document}
\title{Efficient structural relaxation of polycrystalline graphene models}
\author{Federico D'Ambrosio$^1$, Joris Barkema$^2$ and Gerard T. Barkema$^1$}
\address{$^1$Department of Information and Computing Sciences,
Utrecht University, Princetonplein 5,
3584 CC Utrecht, The Netherlands}
\address{$^2$ Informatics Institute, University of Amsterdam, Science Park 904, 1098 XH Amsterdam, The Netherlands}
\begin{abstract}
Large samples of experimentally produced graphene are polycrystalline. For the study of this material, it helps to have realistic computer samples that are also polycrystalline. A common approach to produce such samples in computer simulations is based on the method of Wooten, Winer, and Weaire, originally introduced for the simulation of amorphous silicon. We introduce an \textit{early rejection} variation of their method, applied to graphene, which exploits the local nature of the structural changes to achieve a significant speed-up in the relaxation of the material, without compromising the dynamics. We test it on a $3,200$ atoms sample, obtaining a speedup between one and two orders of magnitude. We also introduce a further variation called \textit{early decision} specifically for relaxing large samples even faster and we test it on two samples of $10,024$ and $20,000$ atoms, obtaining a further speed-up of an order of magnitude. Furthermore, we provide a graphical manipulation tool to remove unwanted artifacts in a sample, such as bond crossings.
\end{abstract}

\maketitle

\section{Introduction}
\begin{figure}[htbp]
\includegraphics[width=\columnwidth]{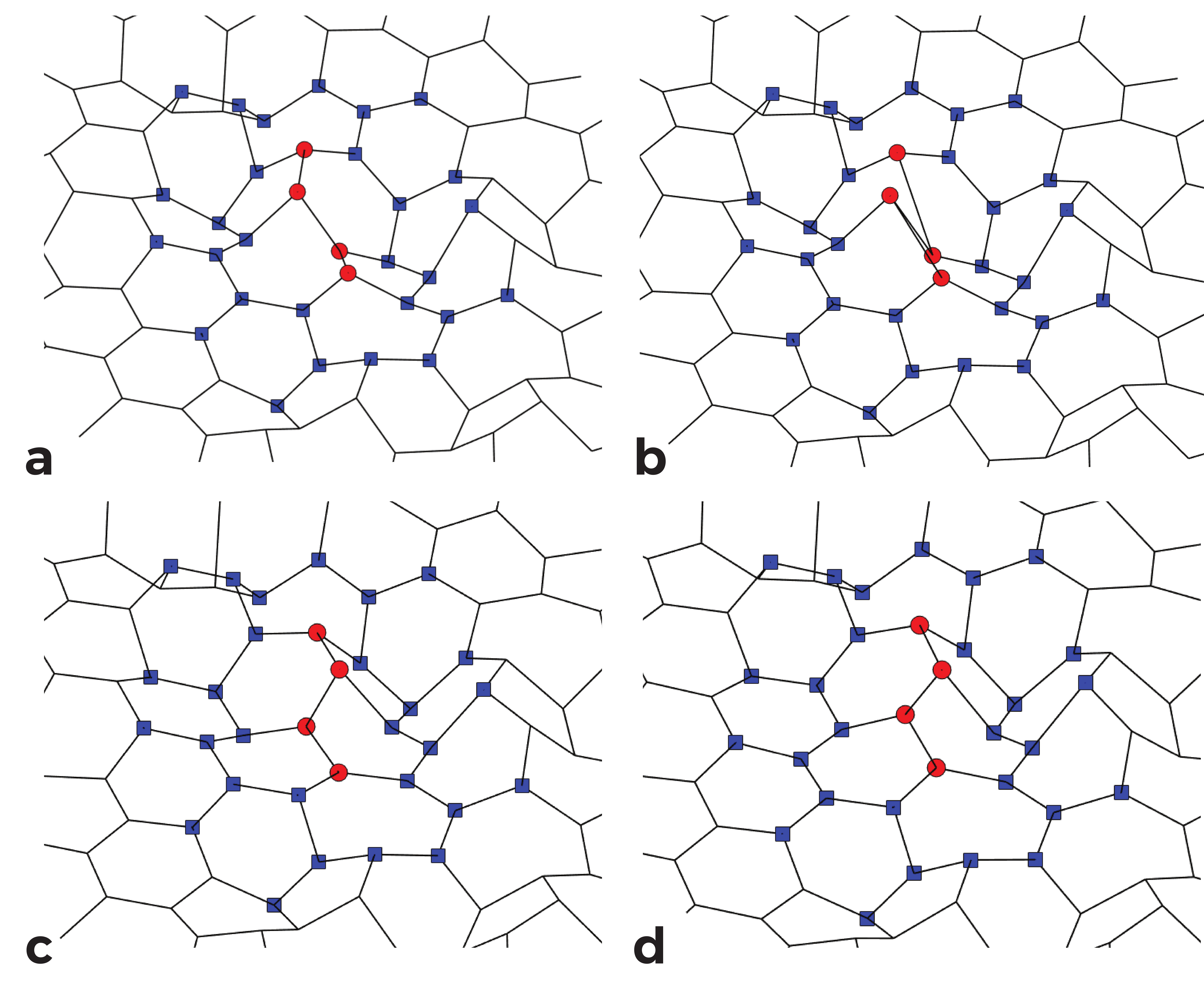}
\caption{Successful bond transposition on a sample of graphene: \textbf{a)} initial configuration \textbf{b)} bond transposition, atoms involved are marked with red dots \textbf{c)} local relaxation, atoms involved are marked with blue squares \textbf{d)} final configuration after global relaxation.}
\label{fig:graphene}
\end{figure}
	
Graphene is a crystal of carbon atoms that form a three-coordinated
honeycomb lattice. It is a material with a large set of exotic properties,
both mechanical and electronic, and it has the particularity
of being a two-dimensional crystal embedded in a three-dimensional
space~\cite{CastroNeto2009,Geim2009,Nair2012,Smith2013,Dolleman2016,Cartamil-Bueno2016,Lee2008,Milowska2013}.
Large samples experimentally produced
are usually polycrystalline, containing
intrinsic~\cite{yazyev2014polycrystalline,Rasool2014,Tison2014},
as well as extrinsic~\cite{Araujo2012} lattice defects. These
defects warrant a thorough study as they both have a
significant detrimental effect on the properties expected from
pristine graphene~\cite{Du2008,BOLOTIN2008351},
and they can also cause new effects that are otherwise
absent~\cite{Lu2013,LIU2015599,Guryel2013,DAmbrosio2019}.

In particular, structural defects are both prominent and common
in graphene~\cite{banhart2010structural}, as they can easily
host lattice defects due to the flexibility of the carbon
atoms in hybridization. Such defects can be frozen in the
sample during the annealing process and have been experimentally
observed~\cite{Hashimoto2004,Meyer2008,PhysRevLett.106.105505}. Their
controlled production in graphene has been explored~\cite{Vicarelli2015}.

Since unsaturated carbon bonds are energetically very
costly~\cite{banhart2010structural}, polycrystalline graphene
samples can be studied with the use of continuous random networks
(CRN) models~\cite{Jain2015}, introduced by Zachariasen almost
90 years ago to represent the lack of symmetry and periodicity in
glasses~\cite{Zachariasen1932}. The rules of this type of model are
quite simple: the only requirement is that each atom is always perfectly
coordinated, i.e. their bonding needs are fully satisfied. Wooten, Winer,
and Weaire (WWW) introduced an explicit algorithm to simulate the evolution of
samples of amorphous Si and Ge, the so-called WWW algorithm that became
the standard for this kind of model~\cite{Wooten1985, Wooten1987}. In
the WWW approach, a configuration consists of a list of the coordinates
of all $N$ atoms, coupled with an explicit list of the bonds between them.

We opted for the empirical potential for polycrystalline graphene
recently proposed by Jain et al.~\cite{Jain2015}:

\begin{equation}
E = \frac{3}{16}\frac{\alpha}{d^2} \sum_{i,j} \left(r_{ij}^2 -
d^2\right)^2\nonumber + \frac{3}{8} \beta d^2 \sum_{j,i,k} \left(
\theta_{j,i,k} - \frac{2 \pi}{3}\right)^2 + \gamma \sum_{i,jkl}
r_{i,jkl}^2 
\end{equation} 

with $r_{ij}$ the distance vector between
the atoms $i$ and $j$, $\theta_{j,i,k}$ the angle centered on the atom
$i$ between the atoms $j$ and $k$, $r_{i,jkl}$ the distance between
the atom $i$ and the plane described by its neighbors $j,k,l$, and
$d = \SI{1.420}{\angstrom}$ the ideal bond-length of graphene. The
other parameters, extracted from DFT calculations~\cite{Jain2015},
are $\alpha = \SI{26.060}{\electronvolt/\angstrom^2}$,
$\beta = \SI{5.511}{\electronvolt/\angstrom^2}$ and $\gamma =
\SI{0.517}{\electronvolt/\angstrom^2}$. The interaction with the substrate
on which the sample lays is simulated by an harmonic confining energy
term in our potential 

\begin{equation}
	E_c = K \; \sum_{i=1} z_i^2
\end{equation} 

where $z_i$ is the z-coordinate of the atom with index $i$
and $K$ a prefactor that is determined empirically in order to constrain
the maximum buckling height to the range of 4-8~$\SI{}{\angstrom}$,
experimentally observed with scanning tunneling microscopy
(TEM)~\cite{Tison2014}.

The process starts with a completely random 2D sample with all atoms
perfectly coordinated, in the case of graphene threefold connected, generated
with the Voronoi diagram algorithm described in~\cite{Jain2015}. In order to generate an initial configuration with $N$ atoms, we place $N/2$ random dots in a 2D square box, which is then surrounded by $8$ copies of itself to implement periodic boundary conditions.  We then compute the $N$ vertices of the Voronoi diagram~\cite{Voronoi1908} of these random dots, which will be replaced by atoms, and connect them along the edges of the diagram to form the bonds by them. This highly-energetic configuration is then carefully relaxed with molecular dynamics.

The
structure of the sample evolves through a series of bond transpositions
involving four connected atoms with two bonds that are broken to
create two new bonds. After each bond transposition, the system is
relaxed; the move is accepted according to the Metropolis acceptance
probability~\cite{Metropolis1953,Hastings1970}:

	\begin{equation}
\label{eq:metropolis-unbiased} 
P(X' | X) = \min \left\{ 1, \; \exp\left[\frac{E(X) - E(X')}{k_b\, T}\right] \right\}
	\end{equation}
	
where $X$ and $X'$ are the configurations of the system respectively
before and after the bond transposition, both the coordinates and the
list of bonds. $k_b$ is the Boltzmann constant, $T$ is the temperature,
and $E(Y)$ the energy of the configuration $Y$ after complete
relaxation. Relaxing the sample, even with an optimized molecular
dynamics algorithm such as the FIRE algorithm~\cite{Bitzek2006},
has a significant computational cost, which is wasted if the bond
transposition is ultimately rejected. As the energy of the sample is
gradually lowered through bond transpositions, the accepted ratio becomes
smaller, often well below one per cent, and almost all computational time
is wasted on proposed bond transpositions that are eventually rejected.

Barkema and Mousseau~\cite{Barkema2000} developed a method for amorphous
silicon, that allows the early rejection of bond transpositions before
completing the relaxation of the sample. It generates a stochastic energy
threshold beforehand, given by \begin{equation} \label{eq:threshold}
E_{t} = E_b - k_b \, T \, \ln(s) \end{equation} with $s$ a random number
between zero and one. In the first ten relaxation steps, the sample is
relaxed only locally up to the third neighbor shell. The energy is assumed
to be harmonic around the minimum, therefore the final energy can be
approximated as proportional to the square of the force 

\begin{equation}
\label{eq:approx}
	E (X') \approx E - c_f |F|^2
\end{equation} 

with $c_f$ an empirically determined constant and $F$ the
force vector. Once we are close enough to the minimum, we can immediately
reject the bond transposition if, at any moment during the relaxation,
$E- c_f |F|^2 > E_{t}$. The efficiency of this method is dependent on the
quality of the assumption Eq.~(\ref{eq:approx}); for amorphous silicon,
the type of model for which it has been developed, this approximation is
generally valid after just a few relaxation steps. 

\begin{figure}[htbp]
\textbf{a)}\includegraphics[width=.47\textwidth]{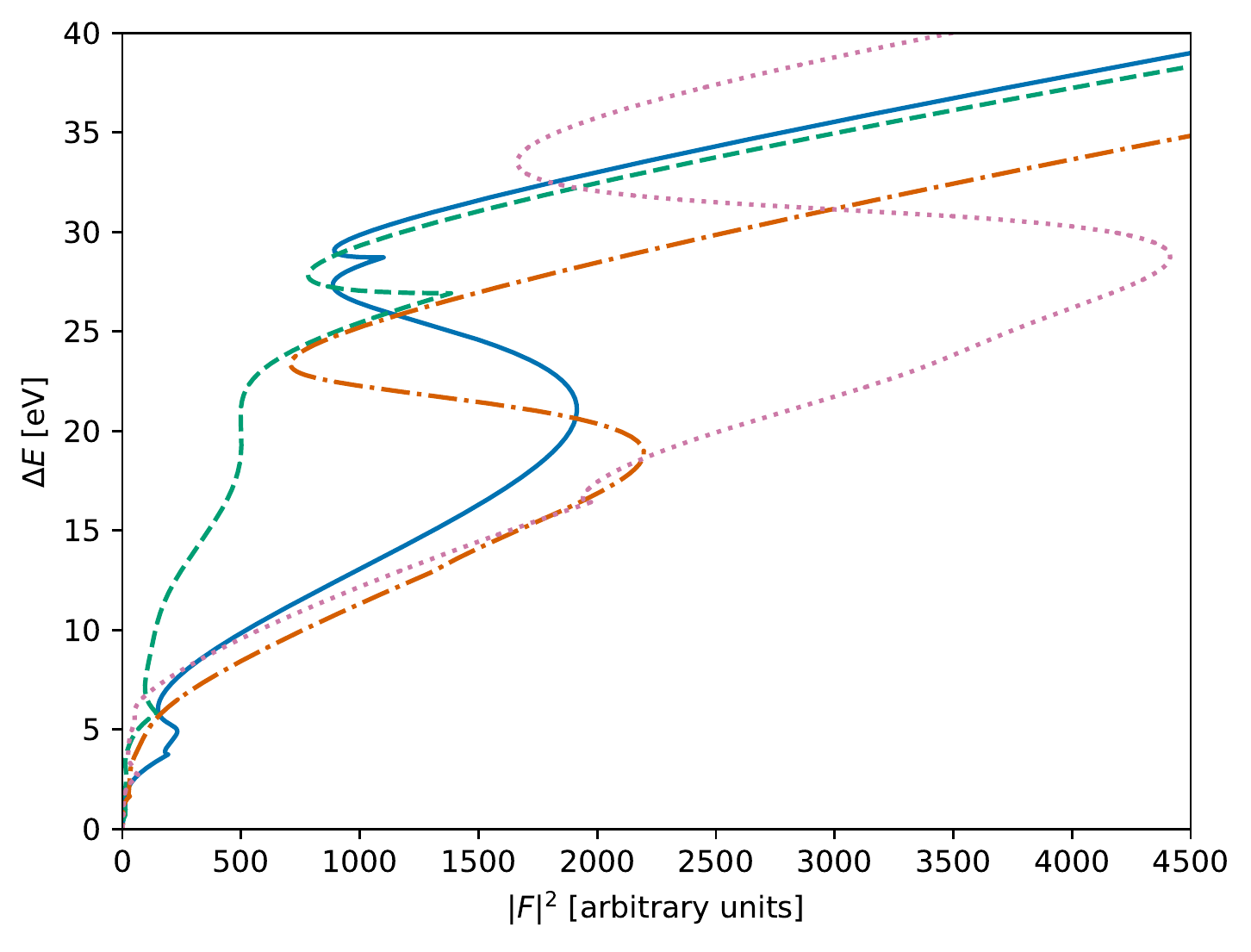}
\textbf{b)}\includegraphics[width=.47\textwidth]{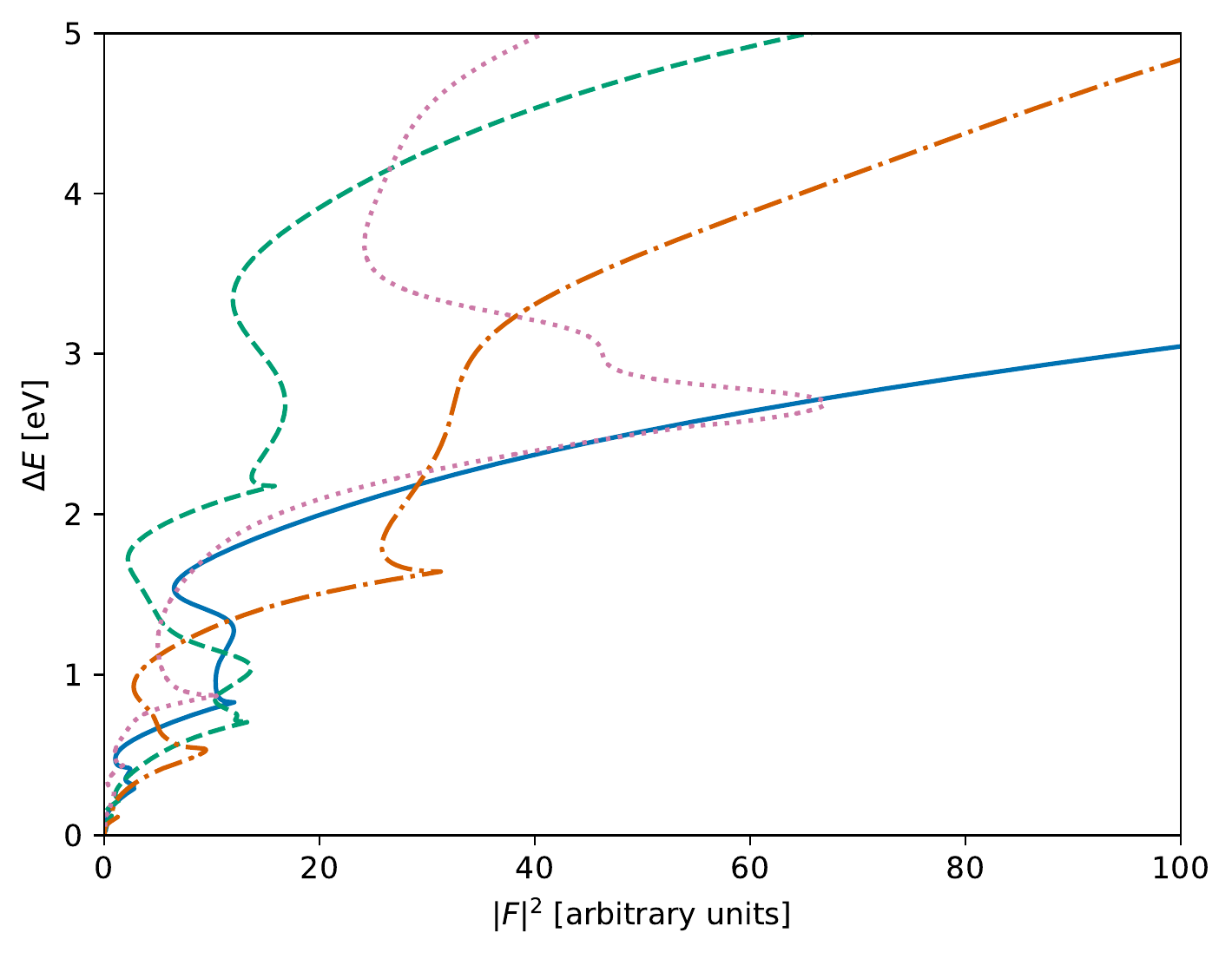}
	\caption{Some typical relaxation trajectories of a 3200 atoms
	sample, farther \textbf{a)} and closer \textbf{b)} to the origin}. Even after thousands of iterations, the approximation
	of Eq.~(\ref{eq:approx}) cannot be applied to this system as it
	fluctuates rapidly in the phase space. $\Delta E$ is the energy
	difference with the relaxed (final) energy, $|F|^2$ the magnitude
	of the forces. \label{fig:cf-BM}
\end{figure}

In theory, this approach could also be applied to polycrystalline
graphene. Unfortunately, the harmonic approximation of
Eq.~(\ref{eq:approx}) is only valid very close to the minimum; as we
show in Figure~\ref{fig:cf-BM}, the trajectory of the system in the
phase-space fluctuates rapidly and erratically during the relaxation,
instead of following the expected linear relation between the excess
energy and squared force magnitude after a certain number of relaxation
steps. Without this approximation, a very costly full relaxation is
necessary after each attempted bond transposition. A different approach
is needed.

In this work, we propose a new method where only the atoms up to
a shortest-path distance $l$ from the atoms involved in the bond
transpositions are initially allowed to relax. The energy of the sample
after this local relaxation is then used to predict the final energy and
immediately reject hopeless bond transpositions, without requiring a
full relaxation. We test this approach on a $3200$ atoms sample, comparing
the performance for different values of $l$. The quality of the results
is also compared to those obtained only through global relaxation. We
further propose a variation of this method for relaxing large samples
and we test it by generating and relaxing a $20,000$ atoms random sample.

\section{Methods}
\begin{figure}[htbp]
\textbf{a)}\includegraphics[width=.47\textwidth]{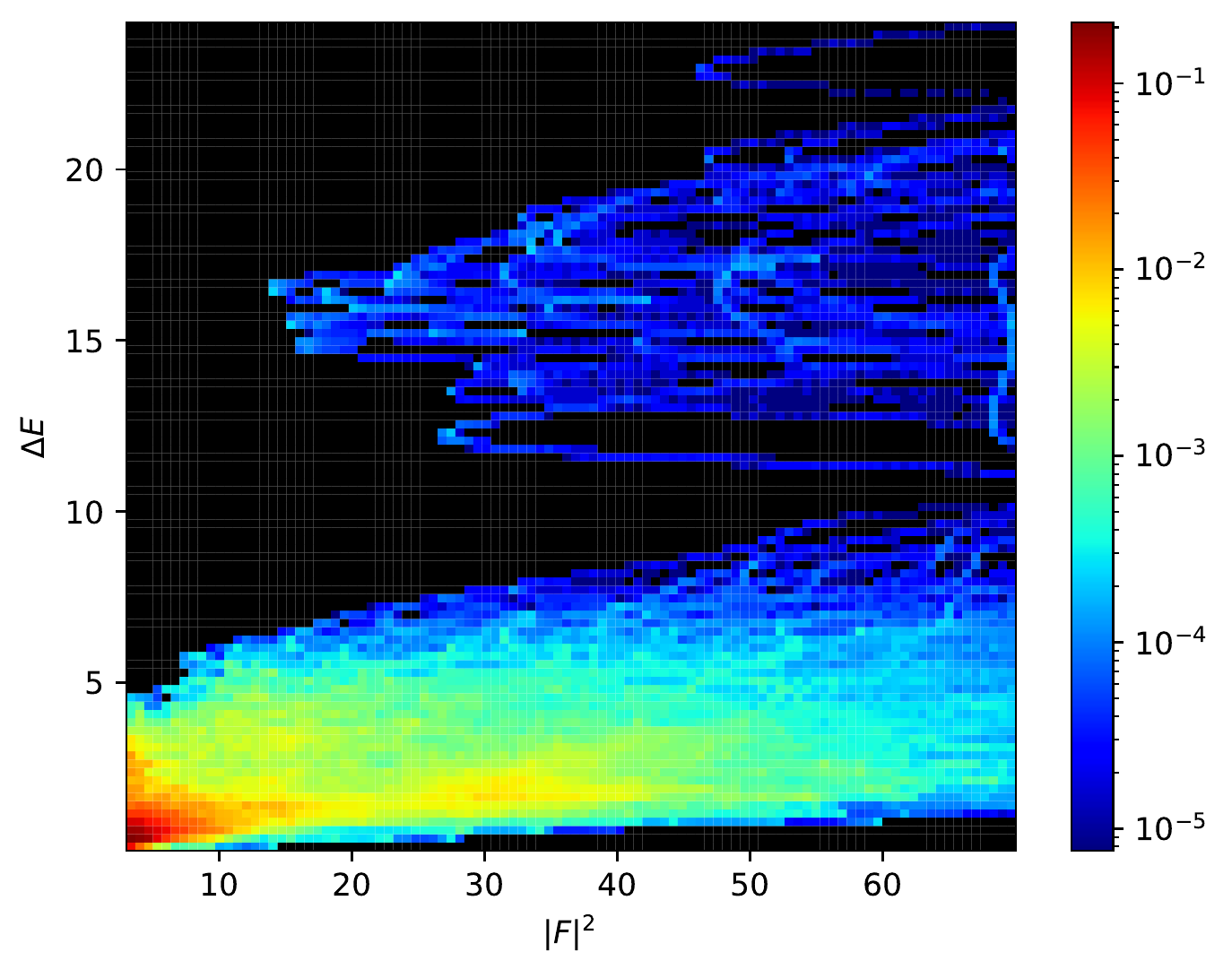}
\textbf{b)}\includegraphics[width=.47\textwidth]{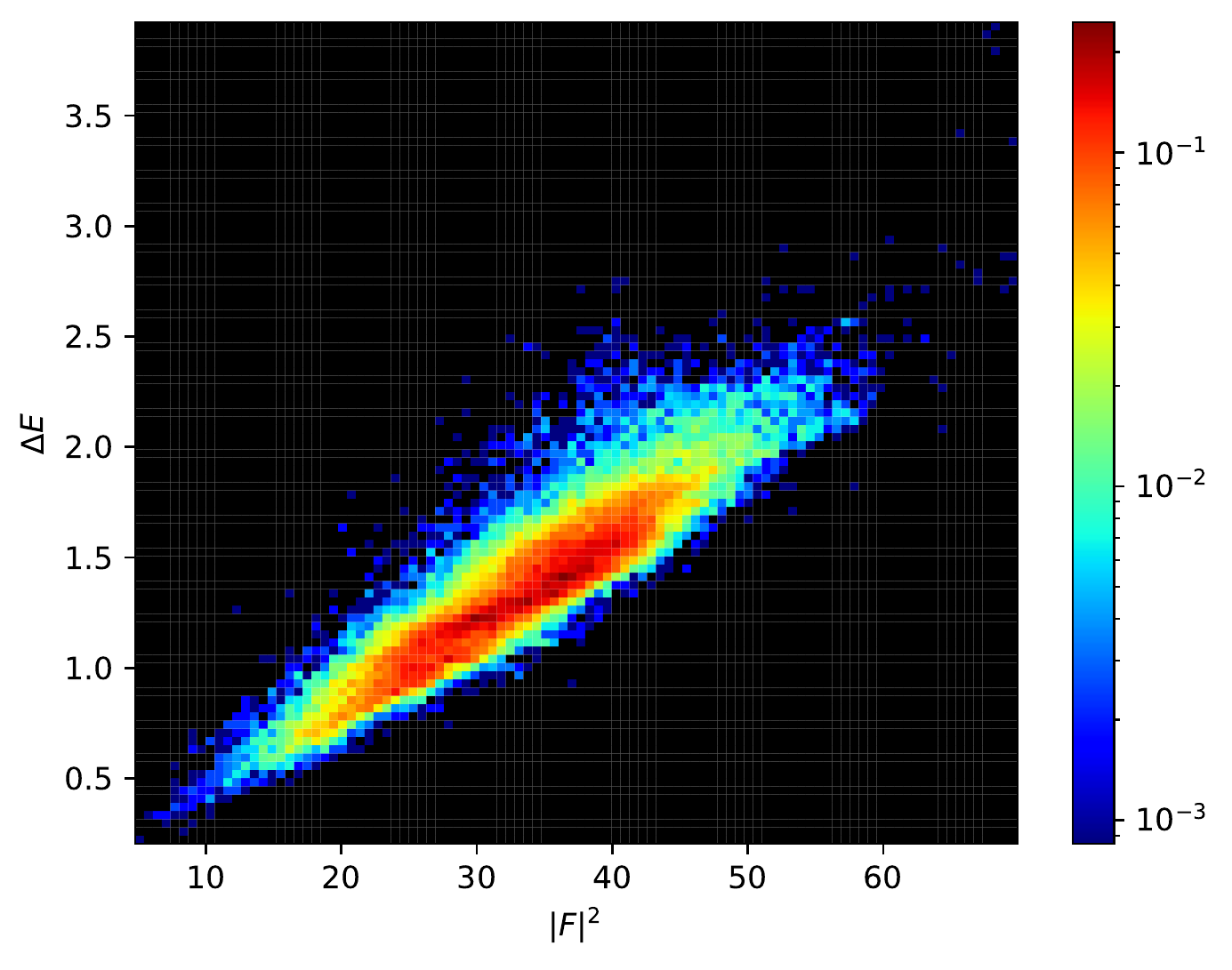}
	\caption{Two dimensional histograms of the energy difference with the relaxed (final) energy $\Delta E$ and the force strength $|F|^2$ over many relaxations of a 3200 atoms sample, from the values assumed during global relaxation, \textbf{a)}, and at the end of local relaxation, \textbf{b)}. Frequency is show in logarithmic scale. We can see that the values assumed at the end of local relaxation follow the harmonic approximation of Eq.~\ref{eq:approx}. } \label{fig:cf-BMvLoc}
\end{figure}

The initial configuration of the sample is a disordered, perfectly
three-fold coordinated, and two-dimensional random network. It is
generated following the procedure described in~\cite{Jain2015}.

The coordinates of the sample are relaxed to an energy minimum with
molecular dynamics, following the FIRE technique~\cite{Bitzek2006}. After
setting a temperature lower than the melting point of graphene, several
bond transpositions are performed until it reaches reasonably low energy
and a realistic configuration. Once this flat sample is sufficiently
relaxed, every atom is placed at a random non-zero distance out of the
two-dimensional plane and allowed to relax to a buckled three-dimensional
configuration.

Our approach to the structural relaxation of graphene can be followed
in Figure~\ref{fig:graphene}. Four consecutive atoms are randomly
selected (Figure~\ref{fig:graphene}a) and the bonds between the first
two and the last two are transposed (Figure~\ref{fig:graphene}b). The
energy threshold is computed from Eq.~(\ref{eq:threshold}) and we
perform a local relaxation around the four atoms involved in the bond
transposition; instead of limiting it to a certain number of relaxation
steps, the atoms up to a shortest-path distance $l$ from the transposed
bonds are allowed to relax completely (Figure~\ref{fig:graphene}c). The
list of atoms involved in the local relaxation is computed after each
attempted bond transposition by iteratively exploring the network,
starting from the four atoms involved in the bond transposition, and
checking against duplicates. After local relaxation, attempted bond
transpositions for which $E_l(X')- c_f |F|^2 > E_{t}$, with $E_l(X')$
the energy of the sample after local relaxation up to distance $l$,
are immediately rejected. In contrast with the method from ~\cite{Barkema2000}, the criterion is applied only once, instead of at each point of the relaxation (with possibly some upper bound on the force strength). As we note in Figure~\ref{fig:cf-BMvLoc}, the force strength after local relaxation (Figure~\ref{fig:cf-BMvLoc}b) is a good estimator for the final energy, especially in comparison to the force strength during the global relaxation (Figure~\ref{fig:cf-BMvLoc}a) , which is used by the method in~\cite{Barkema2000}.

\subsection{Early rejection}
In the \textit{early rejection} approach, the whole sample is otherwise
allowed to relax (Figure~\ref{fig:graphene}d) and, if $E(X') < E_t$,
the bond transposition is finally accepted. As less than one per cent of
proposed moves are accepted in a relaxed sample, we expect the speed-up
to be significant: most are rejected after relaxing a limited number of
degrees of freedom.

The value of $c_f$ is fine-tuned from empirical data collected from
the simulation itself, targeting a higher bound on the rate of false
negatives (i.e. bond transposition that are rejected erroneously),
which in this work was fixed at $2\%$ of the total number of attempts that
should have been accepted. No transposition is accepted without complete
relaxation, regardless of the result of the local minimization; therefore,
no false positive (i.e. a bond transpositions is accepted erroneously)
can be introduced by this technique.

\subsection{Early decision}
While the computational time required for each local relaxation is
constant with regards to the size of the sample, the same cannot be
said for the global relaxation after each tentatively accepted bond
transposition. For large samples, due to the amount of computational
time that this requires, the structural relaxation still grinds almost
to a halt. This is particularly problematic for the initial structural
relaxation of a large random sample, which requires a large number of
bond transpositions to reach a realistic, more relaxed state.

We propose as an alternative for such cases the \textit{early decision}
approach: the decision on whether to reject or accept a bond transposition
after the local relaxation is treated as final, without having to perform
a global relaxation to accept it. The parameter $c_f$ is still fine-tuned
from empirical data but in this case, we opt for the value that best fits
it. After a successful bond transposition, the system will not reach the energy it would have reached with global relaxation and the forces on
atoms outside those involved in the last local relaxation will not go to
zero. To correct for this issue, the energy threshold for accepting a bond
transposition, see Eq.~(\ref{eq:threshold}), will be computed replacing
the current energy of the system ($E_b$) with an estimation of the energy
that our current configuration would reach after a global relaxation
according to Eq.~(\ref{eq:approx}). It can also be useful to set an upper
value for the magnitude of the forces that, when reached, will trigger
a global relaxation that will stop when the magnitude of the forces is
comparable to those that are leftover after a single bond transposition,
to reduce the time spent on these occasional global relaxations.

It must be noted that since we are replacing the energy of the relaxed
system after a bond transposition in Eq. (\ref{eq:metropolis-unbiased})
with an estimate, the \textit{early decision} method does not
guarantee detailed balance, as opposed to the \textit{early rejection}
method. Nevertheless, this method is extremely powerful when performance
is more critical than accuracy, for instance for the structural relaxation
of a very large randomly generated sample when it is still far away from
equilibrium. In these cases, detailed balance is not as critical and a large number of bond transpositions are required to reach a state closer
to the equilibrium.

\subsection{Manipulation tool}
The initial random configuration can incorporate artifacts such as
two bonds crossing each other. While in most cases these defects will
gradually disappear as the sample relaxes to a more ordered configuration,
some artifacts might be particularly resilient and can persist even when
the sample is otherwise sufficiently relaxed. Such defects have to be
removed manually. We have developed a graphical tool called Graphene Editor
\footnote{Available at \url{https://github.com/jorisBarkema/Graphene-Editor}} to facilitate this work. This tool
allows the user to upload and download a sample, explore it visually,
add and remove bonds, move one or more atoms, replace a single atom with
three connected atoms and vice-versa and check the consistency of the
sample over the number of bonds for each atom and bond crossings.

\section{Numerical simulations and results}

\subsection{Early rejection} 
A random sample with $N=3200$ atoms was generated following the procedure
described in the previous section. WWW bond transpositions are performed
until the sample is relaxed to reasonably low energy, approximately
$\SI{625}{\electronvolt}$ (less than $\SI{0.2}{\electronvolt /
atom}$). The values of $c_f$, seen in Table~\ref{tab:cf}, for different
values $l$ of the local relaxation radius are chosen empirically, with the constraint
of keeping the ratio of false negatives (successful bond transpositions
that are nevertheless rejected) over successful bond transpositions
under 2\%, while still rejecting a large part of unsuccessful moves. The quantity $c_f$
is expressed in units of seconds squared over the atomic mass unit
($\SI{}{s^2 \atomicmassunit^{-1}}$). The average number of atoms involved
in the local relaxation for different values of $l$ is also shown in
Table~\ref{tab:cf}.

\begin{table}
\begin{tabular}{c | c | c}
\hline
 $l$ & $c_f~\left[s^2 u^{-1}\right]$ & $\langle N_{loc}\rangle$\\ 
\hline
 1 & $3.21 \cdot 10^{-3}$ & 13\\ 
 2 & $4.63 \cdot 10^{-3}$ & 28\\
 3 & $5.33 \cdot 10^{-3}$ & 53 \\
 4 & $9.08 \cdot 10^{-3}$ & 90 \\
 \hline
\end{tabular}
\caption{Empirically determined values of the harmonic coefficient
$c_f$ and average number of atoms involved in local relaxation $\langle
N_{loc}\rangle$ for different local relaxation distances $l$, in a sample
of size $N=3200$ atoms.}
\label{tab:cf}
\end{table}

\begin{figure}[htbp]
\includegraphics[width=.65\textwidth]{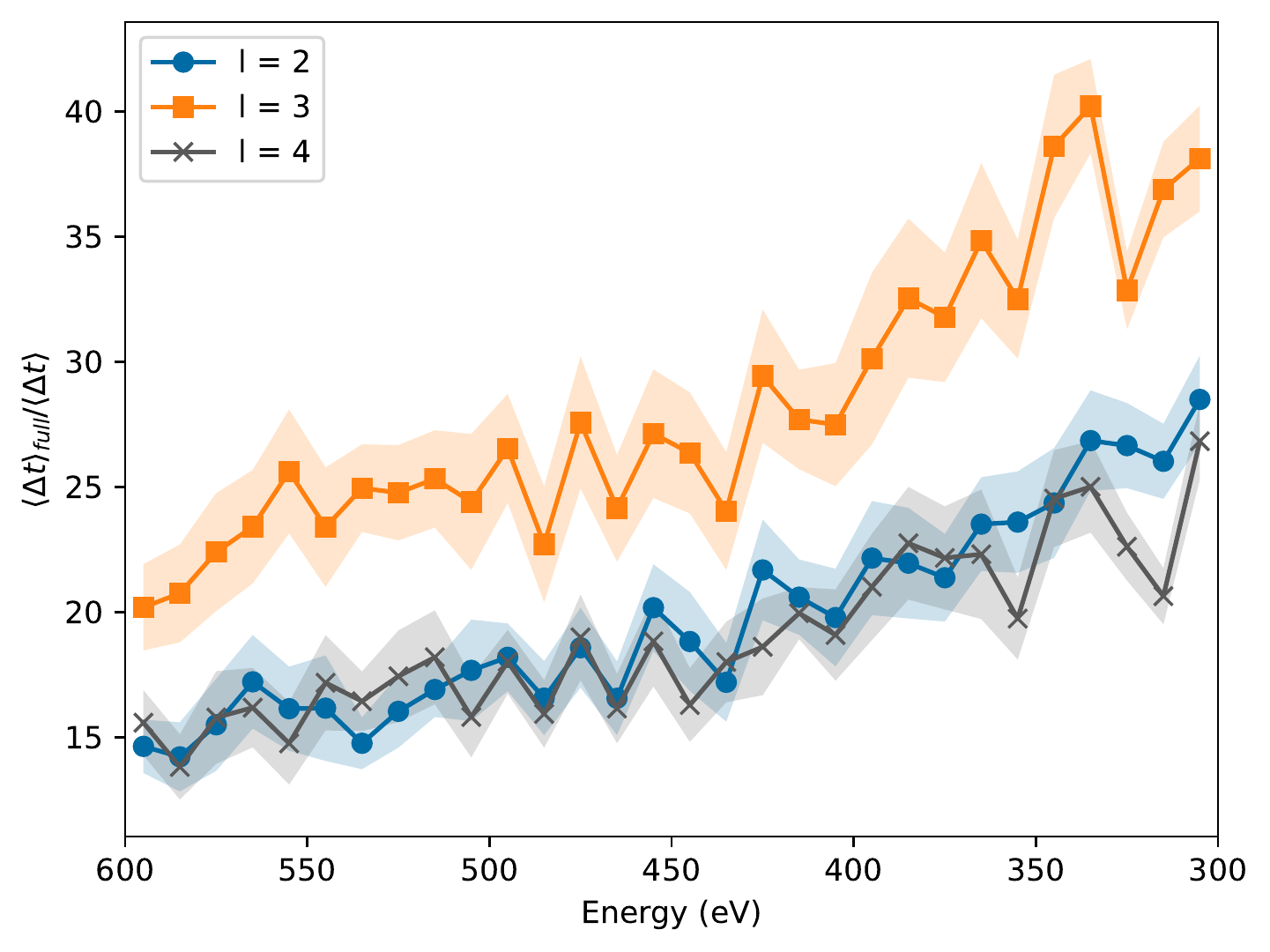}
	\caption{Average speed-up per accepted bond transposition,
	as ratio between CPU time required with full relaxation and
	\textit{early rejection}, for different values of $l$: blue
	dots (2), orange squares (3) and grey crosses (4). The shaded area shows one standard deviation from the average}. The speed
	improvement grows as the sample becomes more crystalline and
	its energy lowers, while best results are obtained for $l=3$,
	with an improvement of a factor between 20 and 40.
	\label{fig:time}
\end{figure}

Starting from the same initial sample, we perform bond transpositions
both using the usual WWW algorithm with full minimization and the
\textit{early rejection} method proposed here, with different values of
$l$. The temperature is set to $T = \SI{3000}{\K}$ for both samples. After each successful bond transposition, we record the energy,
the elapsed time in CPU clocks, and the number of attempts since the
last successful move. The simulation is stopped once the system reaches
a final energy of $E_f = \SI{200}{\electronvolt}$, equivalent to $
\SI{0.0625}{\electronvolt/atom}$. At least ten relaxation cycles are
performed with the \textit{early decision} method (with different values
of $l$) and with complete relaxation after each bond transposition. As
we note in Figure~\ref{fig:time}, the average CPU time per accepted
bond transposition is improved by at least an order of magnitude. The
speed-up grows as the sample grows larger crystalline domains and more
random attempts are necessary per accepted bond transposition. Best
results are obtained for $l=3$, which leads to an efficiency improvement of a factor
between 20 and 40.

\begin{figure}[htbp]
\textbf{a)} \includegraphics[width=.45\textwidth]{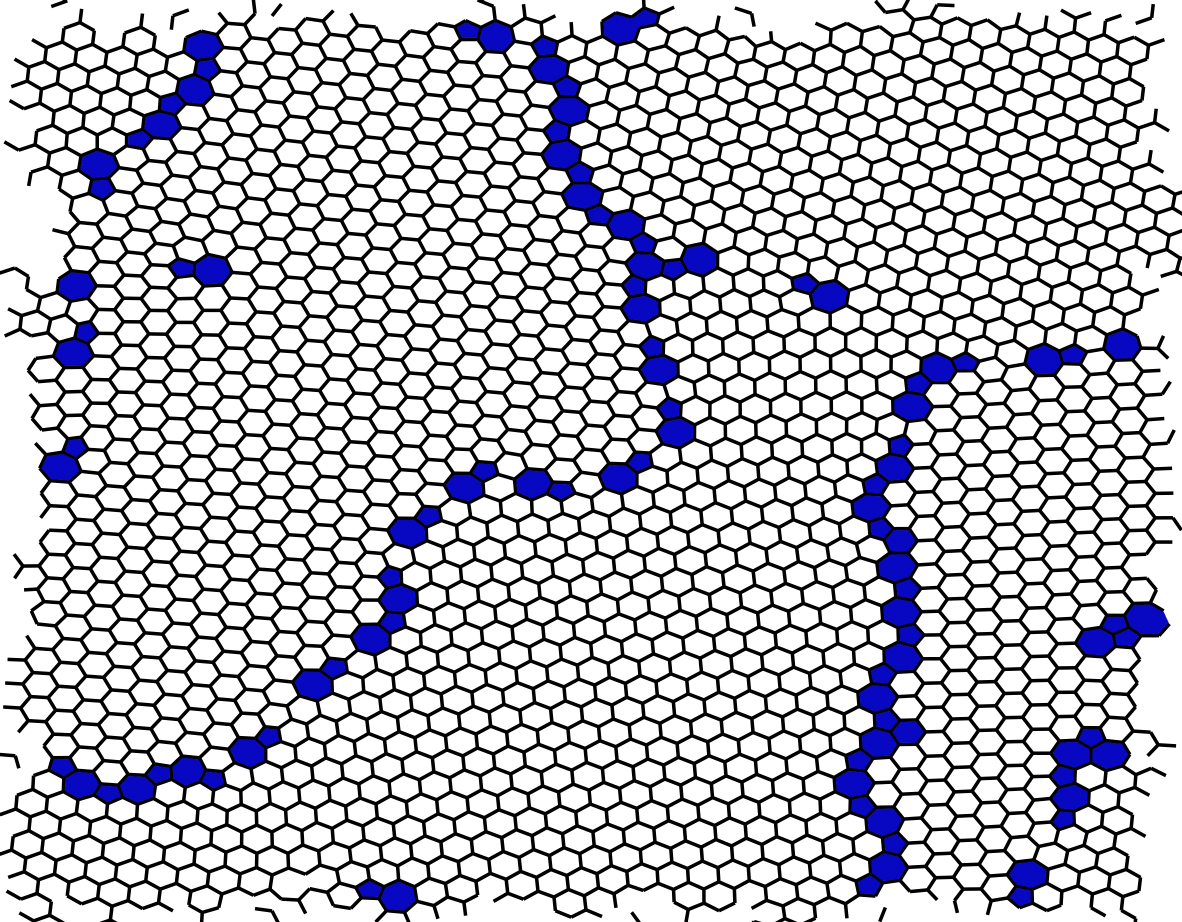}
\textbf{b)} \includegraphics[width=.45\textwidth]{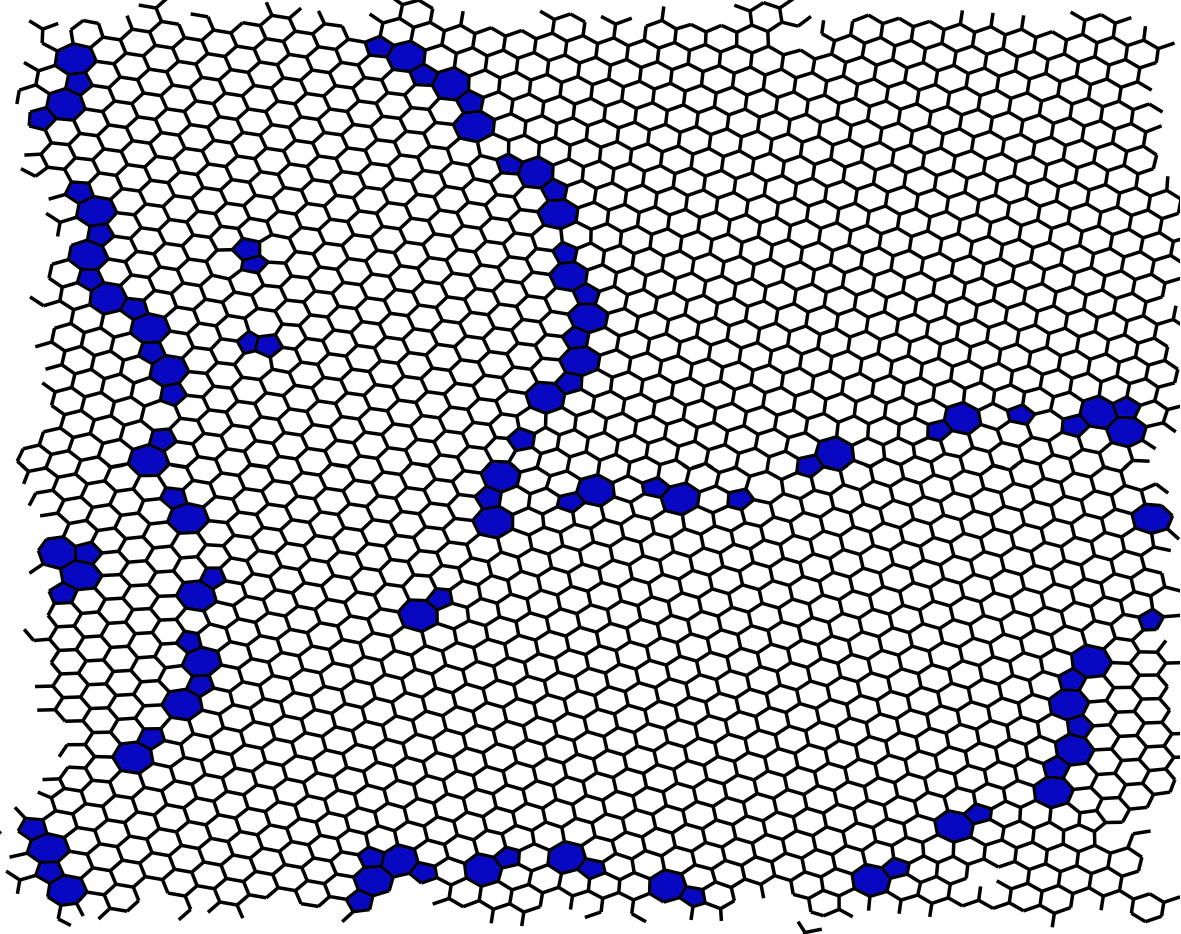}
	\caption{Final configurations of the sample at $E \approx
	\SI{200}{\electronvolt}$ obtained \textbf{a)} through only global
	relaxations \textbf{b)} through our local relaxation method with
	$l=3$. Highlighted in blue are the defective (i.e. non-hexagonal)
	rings. The two samples are qualitatively indistinguishable: same
	level of separation between crystalline domains of similar sizes.}
	\label{fig:end_samples}
\end{figure}

\begin{table}
%%% \tablesize{} %% You can specify the fontsize here, e.g., \tablesize{\footnotesize}. If commented out \small will be used.
\begin{tabular}{l  c  c  c  c}
\toprule
 Atoms &\multicolumn{2}{c}{Full relaxation}&\multicolumn{2}{c}{Early rejection}\\
 \midrule
 Size & \# & \% & \# & \%\\
\midrule
5&56&3.50 &60&3.75\\
6&1489&93.06& 1480&92.50\\
7&54&3.38&60&3.75\\
8&1&$< 0.01$&0&0.00\\
\bottomrule
\end{tabular}
\caption{Ring statistics for the two final configurations of the 3200 atoms sample, relaxed with full relaxation (left) and Early Rejection (right). We note that they both have reached similar statistics, with around $93\%$ of the rings being hexagons, $3 - 4 \%$ heptagons and pentagons, while octagons are too rare at this energy to compare between the two. }
\label{tab:3200-rings}
\end{table}

The \textit{early rejection} method does not alter the amount
of relaxation obtained at the end of the process. As we note in
Figure~\ref{fig:end_samples}, both the level of separation between
crystalline domains, i.e. the degree to which the defects are present on
the borders between them, and the size of the domains are consistent. The ring statistics of the two final configurations, computed with the Ring Statistics Algorithm\footnote{\url{https://github.com/vitroid/CountRings}}~\cite{Matsumoto2007} and reported in Table~\ref{tab:3200-rings}, are also consistent. In
this final configuration, the ratio of false negatives is lower than
0.5\%.

\subsection{Early decision} 
\begin{figure}[htbp]
\includegraphics[width=.65\textwidth]{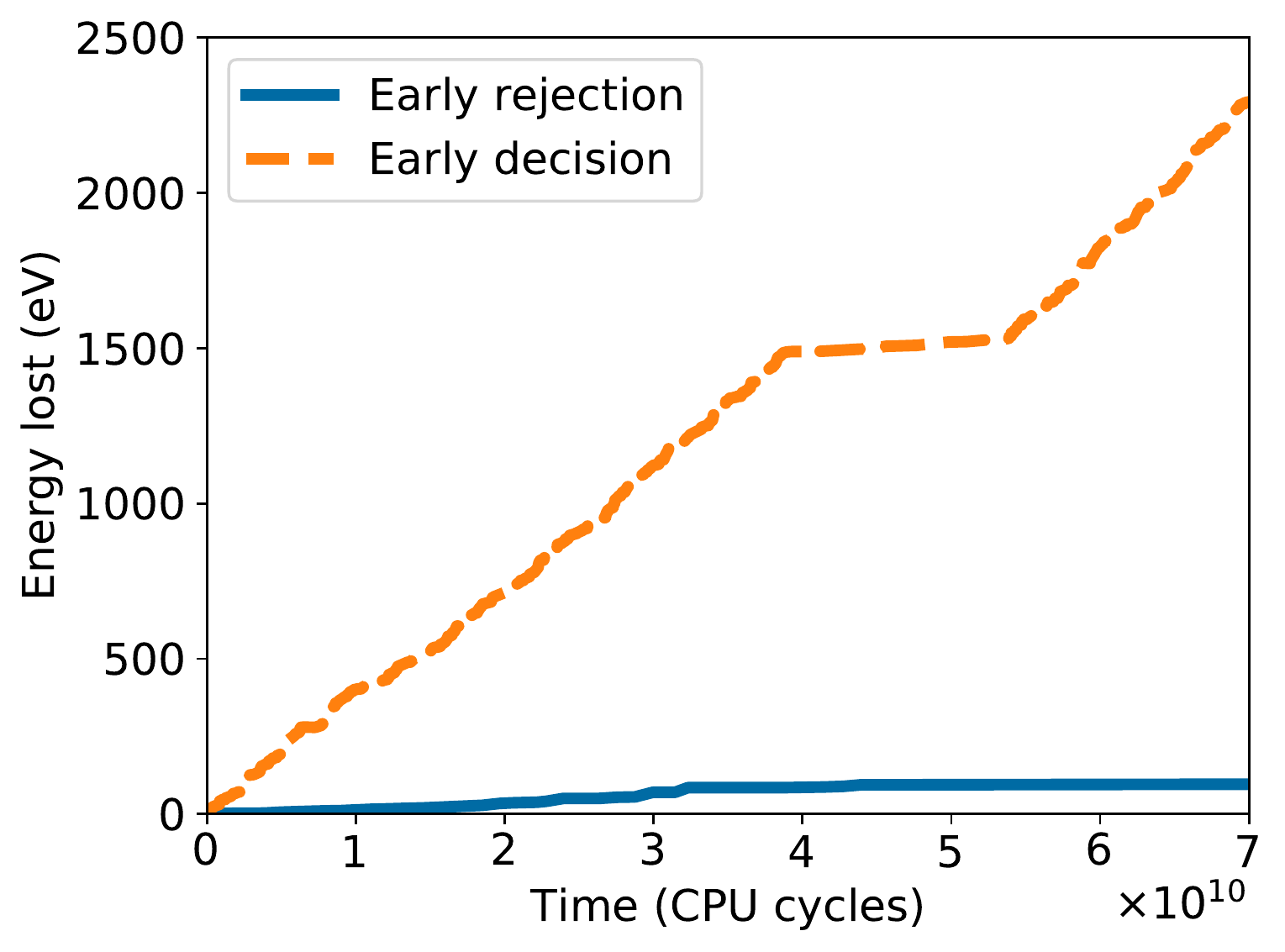}
\caption{Structural relaxation of a large, randomly generated sample
($N = 20,000$), starting from an energy of $\SI{1.15}{\electronvolt
/ atom}$ with \textit{early rejection} (blue solid line) and with
\textit{early decision} (orange dotted line) methods. The \textit{early decision} method
performs significantly faster, reaching a speed-up of a further order of
magnitude. We also notice a plateau around $0.5 \cdot 10^{11}$~CPU cycles
in the \textit{early decision} line where, due to the forces accumulated
from previous bond transpositions, our algorithm was incorrectly rejecting
bond transpositions, slowing the evolution of the sample significantly.}
\label{fig:withandwithoutglobal}
\end{figure}

As we noted in the previous section, while the \textit{early rejection} technique is quite powerful for most samples, it is insufficient for
very large samples; our attempt to relax a very large sample ($N = 20,000$)
could not reach our initial energy target of 1 $eV/\text{atom}$
after more than a month, due to the computational time required by
each global relaxation that takes place at least once per accepted bond
transposition. In the \textit{early decision} method, the decision on
whether to accept a bond transposition or not takes place directly after
performing a local relaxation, based on the estimated relaxed energy of
the sample.

We relaxed with both approaches a randomly generated sample of $20,000$
atoms. We opted again for $l = 3$ for the local relaxation and $c_f$ is set after
fitting the data from \~100 global relaxations. The force magnitude thresholds are set in such a
way that a global relaxation should be triggered each 50-100 successful
bond transpositions and stop when the force magnitude reaches a value
comparable with what is usually left after just one local relaxation.  The temperature is set to $T = \SI{3000}{\K}$.

We initially performed the relaxation on a sample with energy
of $\SI{1.15}{\electronvolt / atom}$. As we can see in Figure
\ref{fig:withandwithoutglobal} the \textit{early decision} approach leads
to a significant speed-up that we estimate to be around one further order
of magnitude. The speed-up factor per bond transposition is stable during
the relaxation at approximately 22. Both methods accept on average a
bond transposition every seven attempts, but the \textit{early decision} method is, as expected, less stable: there can be phases where it is
not able to correctly estimate the correct decision to take. In these
extreme cases, bond transpositions are erroneously rejected and the
evolution of the sample slows down. This is especially the case when
the magnitude of forces accumulated from previous bond transpositions
become significantly large. We can see such a case in the plateau of
the orange dotted line in Figure \ref{fig:withandwithoutglobal} and
it underscores the importance of setting a correct threshold for the
magnitude of forces accumulated before triggering a global relaxation.

\begin{figure}[htbp]
a) \includegraphics[width=.45\textwidth]{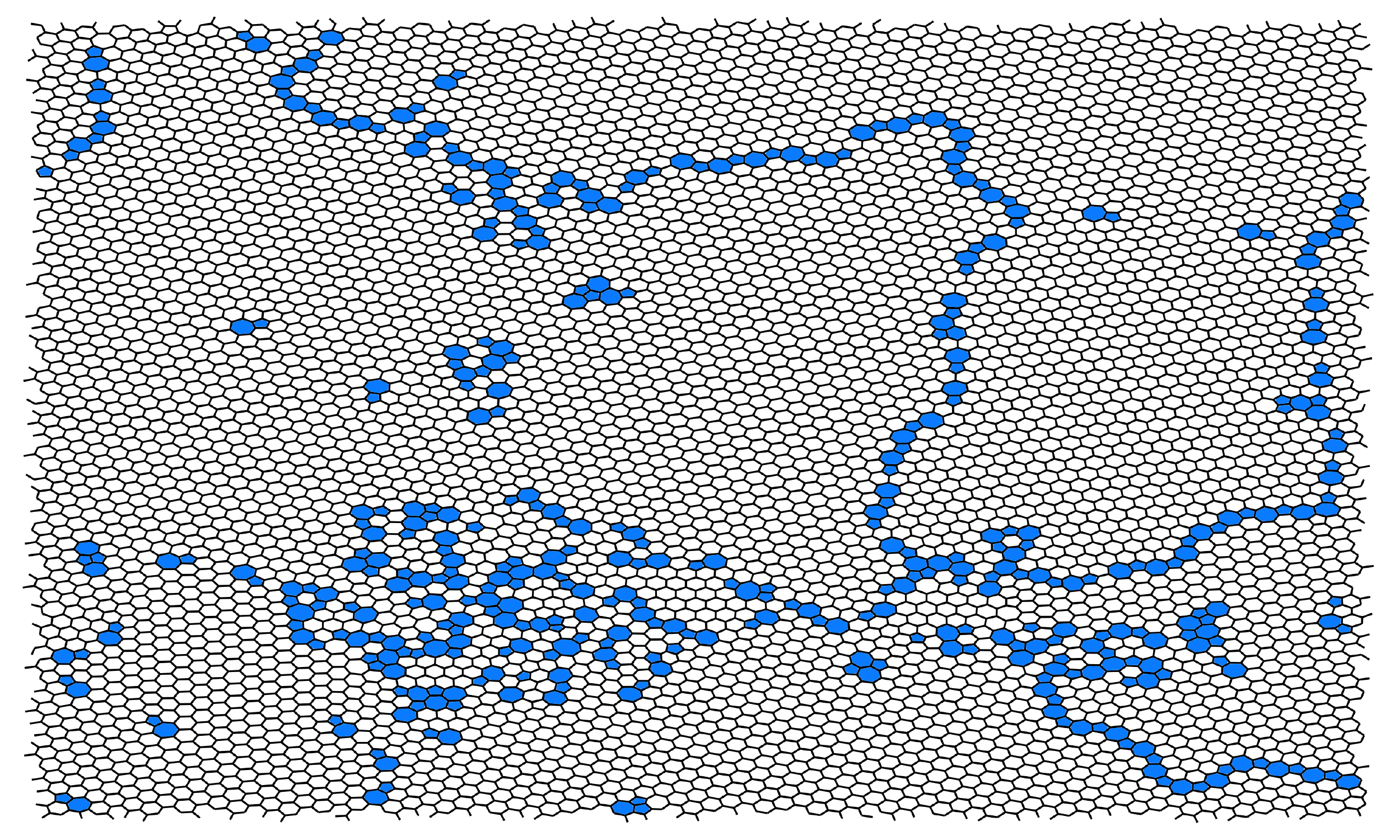}
b) \includegraphics[width=.45\textwidth]{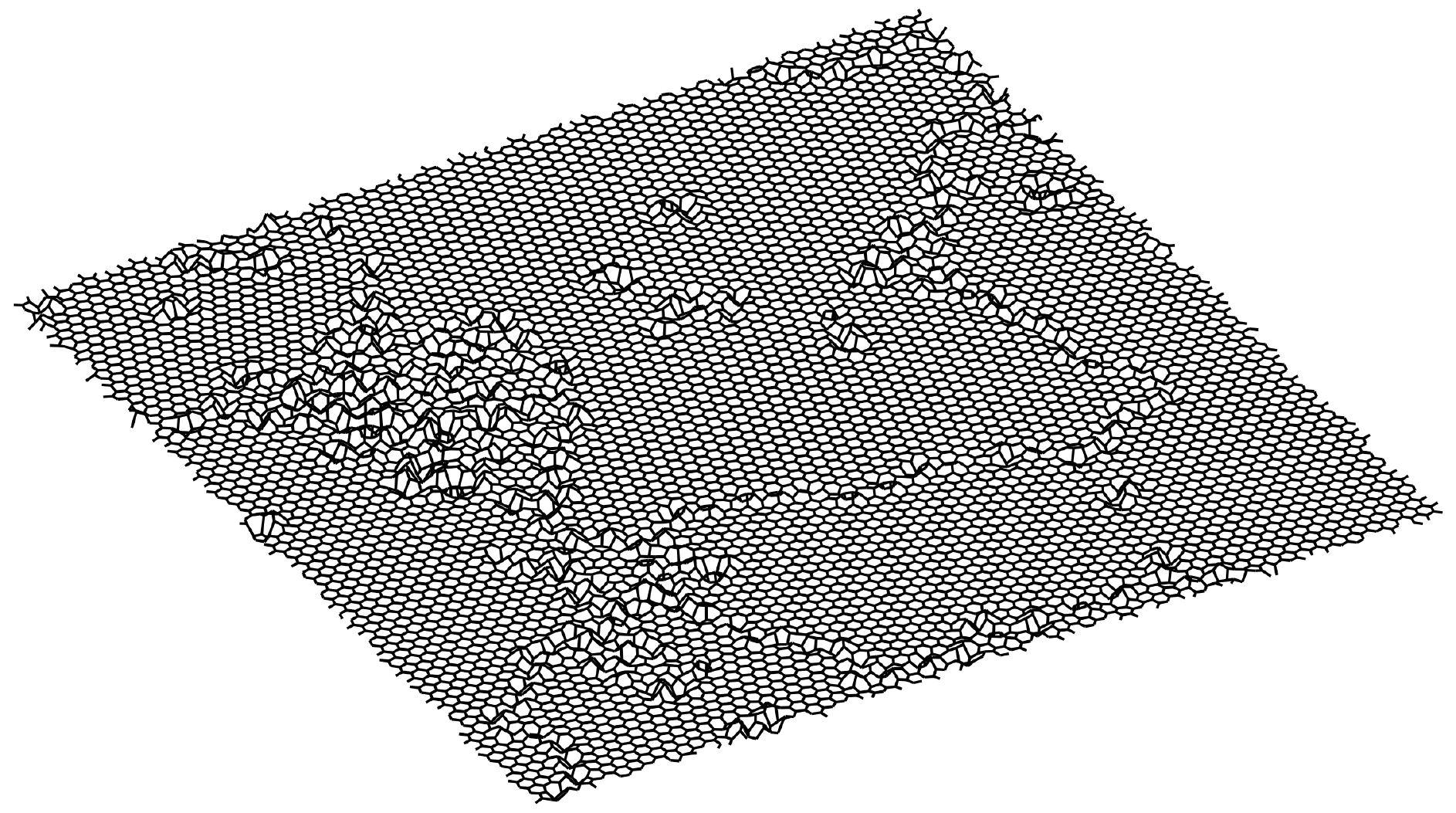}\\
c) \includegraphics[width=.45\textwidth]{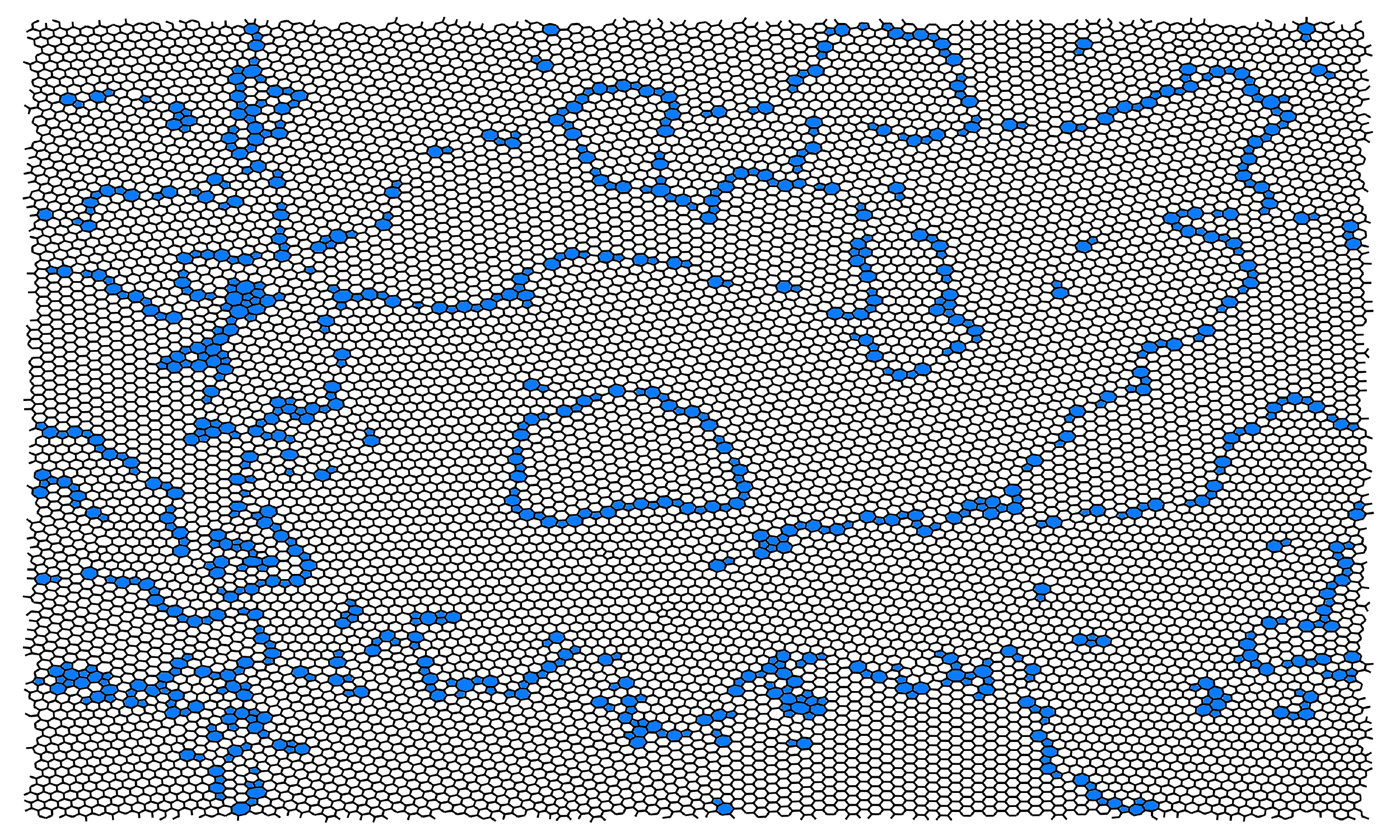}
d) \includegraphics[width=.45\textwidth]{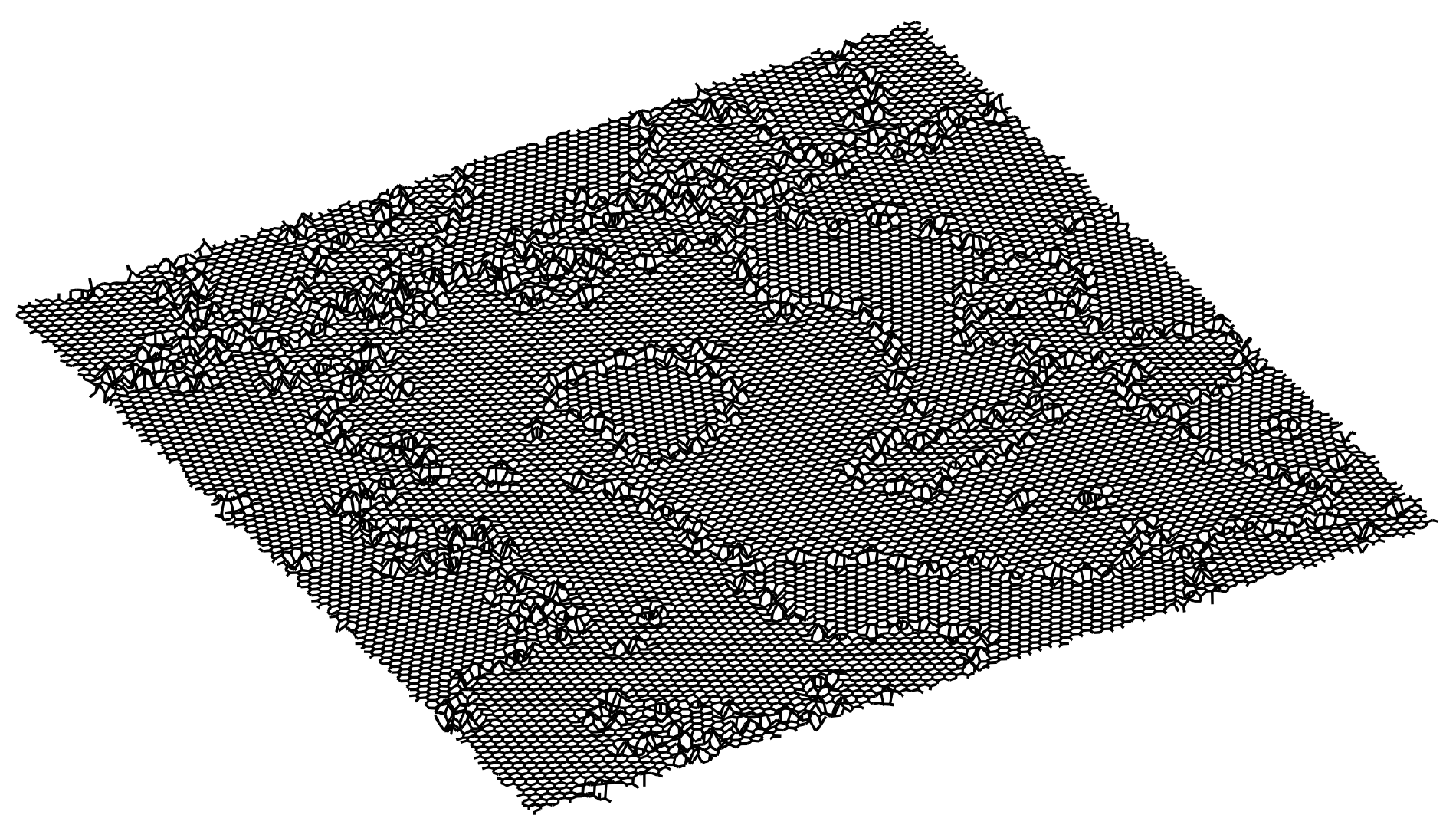}\\
\caption{Final configurations of the $10, 024$ atoms sample in two ($a$) and three ($b$) dimensions and the $20,000$ atoms sample in two ($c$) and three ($d$) dimensions, obtained through \textit{early decision} local relaxation
with $l = 3$. Defects (i.e. non-hexagonal rings) are highlighted in blue in the two-dimensional plots and clearly visible due to the buckling in the three-dimensional plots. The two samples are qualitatively very similar to those of Figure~\ref{fig:end_samples}, with large domains surrounded by defects.}
\label{fig:10and20k}
\end{figure}

\begin{table}
\begin{tabular}{l  c  c  c  c}
 Atoms &\multicolumn{2}{c}{$10,024$}&\multicolumn{2}{c}{$20,000$ }\\
 Size & \# & \% & \# & \%\\
\hline
5&231&4.6 &487&4.87\\
6&4554&90.86& 9043&90.43\\
7&223&4.45&453&4.53\\
8&4&0.08&17&0.17\\
 \hline
 \end{tabular}
\caption{Ring statistics for the two large samples of $10,024$ and $20,000$ atoms. We note that they both have reached similar statistics, with over $90\%$ of the rings being hexagons, $4 - 5 \%$ heptagons and pentagons and less than $0.2\%$ octagons.}
\label{tab:rings}
\end{table}

Finally, we relaxed the $20,000$ atoms sample and another sample of $10,024$ atoms down to $\SI{1488.05}{\electronvolt}$ ($\SI{0.074}{\electronvolt / atom}$) and $\SI{695.51}{\electronvolt}$ ($\SI{0.066}{\electronvolt / atom}$) respectively. The temperature is initially set at $\SI{3000}{\K}$ and then gradually reduced, in order to reach lower energies. The resulting samples, as we note in Figure~\ref{fig:10and20k}, present large crystalline domains with defects accumulating on their boundaries, similarly to~\ref{fig:end_samples}. As we note in Table~\ref{tab:rings}, both samples have reached similar ring statistics, with less than $10\%$ of defected rings and only a handful (less than $0.2\%$) defected by more than one atom (i.e. octagons). The ring statistics is computed with the Ring Statistics Algorithm~\cite{Matsumoto2007}.

All the samples presented in this paper are available online\footnote{\url{https://github.com/federicodambrosio/graphene-samples}}.

\section{Discussion and outlook}
In summary, we introduced two techniques that, through local relaxation, can estimate the success of bond transpositions reducing or
eliminating the need for relaxing the entire sample, which is extremely
time-consuming. Both techniques significantly reduce the computational
time required per accepted bond transposition: the \textit{early
rejection} method by immediately rejecting, without a global relaxation,
hopeless attempts; the \textit{early decision} method avoids global
relaxations entirely, relying on the estimate of the energy of the
relaxed sample.

The \textit{early rejection} technique should be preferred for
average-sized samples, especially if already well-relaxed since it gives
an already significant speed-up while it guarantees that the dynamics are
not compromised. Furthermore, its accuracy also improves as the energy
of the sample is reduced. The \textit{early decision} technique leads
to an even larger speed-up but does allow for attempts to be erroneously
rejected and should therefore be used when performance is a priority
above accuracy, for instance when the sample is still very far from equilibrium and detailed balance is less critical. Since thousands of
bond transpositions are required to reduce the energy of a few hundreds
of electron volt, the cumulative speed-up obtained through either of
these techniques can easily reach multiple orders of magnitude. These
techniques open up the possibility of generating larger random samples
with ordinary computers in an affordable amount of time. 

Finally, our \textit{manipulation tool} Graphene Editor makes those small manipulations
that are often necessary as simple and quick as they can be.

\section{Author contribution statement}
F.D. and G.T.B. conceived the project and contributed to the theoretical analysis. F.D. implemented the simulations and wrote the manuscript. J.B. developed the Graphene Editor tool.

\bibliographystyle{apsrev4-1}
\bibliography{references-clean}

\end{document}